\documentclass[ nofootinbib,aps, twocolumn]{revtex4}
\usepackage{amsmath,amssymb,graphicx}
\usepackage{epsf}
\usepackage{epstopdf}
\usepackage {amssymb}
\usepackage{color}
\newcommand{\nc}{\newcommand}
\nc{\ba}{\begin{eqnarray}}
\nc{\ea}{\end{eqnarray}}
\newcommand\be{\begin{equation}}
\newcommand\ee{\end{equation}}

\newcommand{\calP}{{\cal{P}}}
\newcommand{\calS}{{\cal{S}}}

\nc{\x}{{\bf{x}}}
\nc{\A}{{\cal A}_{eff}}
\nc{\Cl}{{C_\ell}}

\begin{document}
\title{Expected dipole asymmetry in CMB polarization}

\author{M.  H. Namjoo$^{1}$}

\author{A. A.  Abolhasani$^{2,3}$}

\author{H. Assadullahi$^{ 4, 5}$}

\author{S. Baghram$^{2}$}

\author{H. Firouzjahi$^{6}$}

\author{D. Wands$^{4}$}


\affiliation{$^1$ Department of Physics, The University of Texas at Dallas, Richardson, TX 75083, USA}

\affiliation{ $^2$ Physics department, Sharif University of Technology, Tehran, Iran}

\affiliation{$^3$School of Physics, Institute for Research in
Fundamental Sciences (IPM),
P.~O.~Box 19395-5531,
Tehran, Iran}

\affiliation{$^4$ Institute of Cosmology and Gravitation, University of Portsmouth, Dennis Sciama Building, Burnaby Road, Portsmouth PO1 3FX, United Kingdom}
\affiliation{$^5$ School of Earth and Environmental Sciences, University of Portsmouth, Burnaby Building, Burnaby Road, Portsmouth PO1 3QL, United Kingdom}
\affiliation{$^6$School of Astronomy, Institute for Research in
Fundamental Sciences (IPM),
P.~O.~Box 19395-5531,
Tehran, Iran}

\begin{abstract}
We explore the hemispherical asymmetry predicted in cosmic microwave background polarization when there is an asymmetry in temperature anisotropies due to primordial perturbations. We consider the cases of asymmetries due to adiabatic and isocurvature modes, and tensor perturbations. We show that the asymmetry in the TE, EE and/or BB correlations can be substantially larger than those in the TT power spectrum in certain cases. The relative asymmetry in the different cross-correlations, as well as the angular scale dependence, can in principle distinguish between different origins for the asymmetry.
\end{abstract}

\maketitle

\section{Introduction}

Studies of the cosmic microwave background (CMB) fluctuations have become a cornerstone of 
modern cosmology. The standard model of cosmology, known as $\Lambda$CDM,
with initial conditions set up during an early period of inflation is supported by detailed CMB 
observations, as demonstrated by Planck  \cite{Ade:2013zuv, Ade:2013uln} and WMAP \cite{Komatsu:2010fb, Hinshaw:2012aka} experiments.  However, despite overall confirmation of the basic picture, there are some indications of unexpected features in the CMB sky. In particular the Planck collaboration reported a hemispherical power asymmetry in the large-scale temperature anisotropies \cite{Ade:2013nlj}, which was also observed in WMAP results \cite{Eriksen:2003db, Hansen:2004vq, Eriksen:2007pc, Hansen:2008ym, Hoftuft:2009rq}. 

A simple parameterization of hemispherical power asymmetry in temperature anisotropy $\delta T$ is \cite{Gordon:2006ag}
\ba
\label{dipole-eq0}
{\delta T}(\hat{ {\bf n} })=\overline{\delta T}({ \bf \hat{n}}) \big( 1+A \, {\bf  \hat p.\hat n}  \big) 
\ea
in which  $\overline{\delta T}(\bf{\hat{n}})$ is the isotropic part of the temperature fluctuations, $\bf{\hat{n}}$ is direction of the observation, $A$ is the amplitude of the hemispherical asymmetry and ${\bf\hat{p} }$ is the preferred direction in CMB temperature map.  The Planck collaboration  has found the amplitude $A=0.073\pm0.010$ and the preferred direction ${\bf \hat{p}} =  (217.5\pm15.4 , -20.2\pm 15.1)$ in galactic coordinates for large angular scales, corresponding to low multipoles $\ell=2-64$ \cite{Ade:2013nlj}.  This apparent detection 
has generated significant interest; for some proposed explanations for the origin of asymmetry see \cite{Donoghue:2007ze, kam1,kam2,kam3,Assadullahi:2014pya,pesky, lyth, Namjoo2013, Abolhasani2013, Namjoo:2014nra, 
jm1, jm2, McDonald:2014lea, McDonald:2014kia, 
Wang2013, Mazumdar:2013yta, Lou2013,Liddle2013, Amico2013, Cai:2013gma, Rath2013, Kohri2013, Kanno2013, Firouzjahi:2014mwa, Jazayeri:2014nya, Baghram:2014nha, Liu:2013iha, Fernandez-Cobos:2013fda, Chang2013,Wang2013a, Rath:2014cka, Lyth:2014mga} and for the related data analysis of CMB and large-scale structure constraints on dipole asymmetry see \cite{Hirata:2009ar, Akrami:2014eta, Flender:2013jja, Notari:2013iva, Quartin:2014yaa, Adhikari:2014mua}. 

The observed amplitude of the dipole 
has a non-trivial scale-dependence such that on small scales the allowed amplitude of the dipole asymmetry becomes significantly smaller \cite{Hirata:2009ar, Akrami:2014eta, Flender:2013jja, Adhikari:2014mua}, however see also \cite{Quartin:2014yaa}. 
Using quasar data it was concluded in \cite{Hirata:2009ar} that $A<0.012$ at  $2\sigma$ level on Mpc scales.  
Also, recent analysis with Planck data implies that there is no sign of dipole asymmetry on angular scales
 $\ell > 600$ \cite{Flender:2013jja, Adhikari:2014mua}. 
 
Theoretically, there is no compelling model of inflation to address the origin of hemispherical asymmetry with the required scale-dependence to fit the observational data. 
However, one generic picture which has emerged is that in order to generate a large observable dipole asymmetry, by means of non-linear correlations of CMB scale modes with a very long wavelength superhorizon mode, one has to go beyond a single field slow-roll inflation model. As demonstrated  in  \cite{Namjoo2013, Abolhasani2013, Namjoo:2014nra}, see also \cite{lyth, Lyth:2014mga}, the amplitude of dipole asymmetry can be related to the level of non-Gaussianity in the squeezed limit. Therefore, to generate a large observable dipole asymmetry one has to break the Maldacena's consistency condition in simple single field models of inflation \cite{Maldacena:2002vr}. For example, models of inflation with multiple fields \cite{Wang2013, Mazumdar:2013yta, Namjoo2013} or with non-vacuum initial conditions \cite{Firouzjahi:2014mwa} can in principle provide a platform to generate 
a large dipole asymmetry due to the correlation between a very long wavelength super-horizon mode and CMB scale modes. 
 
The Planck team is expected to soon present its full temperature and polarization maps.  Assuming that the temperature map shows a dipole asymmetry of primordial origin in the form of Eq. (\ref{dipole-eq0}), one naturally anticipates that a corresponding signal should also be present, and in principle observable, in the polarization maps and their correlations.  The expected form and amplitude of this asymmetry is the main goal of this paper. We investigate how a dipole asymmetry in the primordial fluctuations can generate dipole asymmetries in $TT, EE, TE$ and $BB$ correlations. We consider primordial dipole asymmetries generated from different types of modes  (a) adiabatic, (b) matter isocurvature and (c) tensor modes, and then study their effects on $TT, TE, EE$ and $BB$ correlations.  We show  that each mode has distinct and in principle distinguishable effects, on polarization correlations. That is, the scale-dependence as well as the amplitude of asymmetry will be different for each mode. In addition, we also investigate the cases in which two independent modes, i.e. adiabatic+isocurvature, adiabatic+tensor and isocurvature+tensor, jointly generate the CMB asymmetry. 

The methods employed in this work are primarily phenomenological with our motivation being to explore the main contributions from each type of mode (adiabatic, matter isocurvature and tensor) in temperature and polarization asymmetries. We can show that for various cases the numerical results obtained here can be understood analytically.  We do not provide here the specific analytic results behind the current results, which are left for future work \cite{progress}.
We do not rely on particular model as the source of dipole asymmetry, although as a good example one may refer to mechanism of dipole asymmetry generated from long mode modulation \cite{pesky, lyth, Namjoo2013, Abolhasani2013, Namjoo:2014nra}. The advantage of our approach 
here is that the predictions are model independent. However, in the case of asymmetry generated from more than one mode we assume the same preferred direction for all independent modes. This  implies that the asymmetries in different modes have a common origin though still it is not limited to a specific model. 

\section{General formalism}
\label{SecII}

In this section, we present a general formalism for the scale-dependent asymmetry in our setup.
We will consider three types of asymmetric primordial perturbations (a) adiabatic, (b) matter isocurvature and (c) tensor modes as the modulating sources of CMB dipole asymmetry. We follow a phenomenological approach and do not specify the origin of these asymmetries. We assume that the modes have a common origin, hence are all in the same direction but have independent amplitudes of asymmetry. The isocurvature and the tensor modes both have a scale-dependent signature in the CMB as they rapidly decay for $\ell>100$. In contrast, the adiabatic mode does not decay as quickly on small scales. Hence, in order to generate the observed scale-dependent dipole asymmetry in temperature anisotropies, which falls off for higher values of $\ell$, we  assume a scale-dependent primordial asymmetry for the adiabatic mode and let it to decay away for scales corresponding to $\ell \gtrsim 64$. 

Let us start with the adiabatic mode in which the curvature perturbation is parameterized as
\ba
\zeta_k = \zeta_k^0 (1+A_\zeta \, f(k)\, {\bf  \hat p.\hat n} )
\ea
in which $\zeta^0$ is the isotropic curvature perturbation and $\bf{\hat{p}}$ and $\bf{\hat{n}}$ are the preferred direction and the line of sight, respectively. Here $f(k)$ controls the scale-dependence of the asymmetry and $A_\zeta$ is the overall amplitude of the asymmetry in the curvature perturbation. In the case of  scale-independent asymmetry, $f(k)=1$, we require $A_\zeta=0.07$ to fit the current data at low multipoles, though strict scale independence is not compatible with observations for $\ell > 600$ as discussed in the introduction.   

To obtain the effective amplitude of the asymmetry in the more general case, we compare the results for two extreme power spectra. The first case is when the line of sight is perpendicular 
to the preferred direction, $ {\bf  \hat p.\hat n} =0$, so there is no correction to power spectrum denoted by $\calP^0$. The second case is when the observer looks along the direction of 
the asymmetry, $ {\bf  \hat p.\hat n} =1$,  in which case the asymmetric corrections to the power spectrum are maximal  $\calP (k) = \calP^0(k) + \Delta \calP(k)$, where
\ba
\Delta \calP(k)  =  \calP^0(k) \big[\, (1+A_\zeta \, f(k))^2 -1 \big] \simeq 2 \calP^0(k) A_\zeta \, f(k) \,.
\ea
From the above primordial power spectra, one can easily run CAMB to obtain the $TT$, $TE$ and $EE$ correlations from $\calP^0$ which we denote by $\Cl^{\zeta_0}$.
For linear perturbations one can obtain $\Cl^\zeta$, i.e., the $\Cl$ obtained from the primordial power $\calP^0(k) + \Delta \calP(k) $, from a linear combination
\ba
\Cl^{\zeta} = \Cl^{\zeta_0} +2 A_\zeta \Cl^{\Delta \zeta } \,,
\ea
in which $\Cl^{\Delta \zeta } $ is obtained from the residual power spectrum $\Delta \calP(k) $ 
but we kept the amplitude $A_\zeta$ outside of $\Cl$ which will be useful for future calculations. An important assumption for the last equation is the linearity of all events from primordial perturbations to the observed $\Cl$'s. This is the case if one neglects the non-linear effects such as lensing \cite{Lewis:2006fu}. Since we are only interested in the range of scales $2 <\ell <64$, the latter assumption is very good for $TT$, $TE$ and $EE$ correlations while it induces nearly $10\%$ error around $\ell \sim 60$ for $BB$ correlation. As a result, the effect of lensing on the scale dependent asymmetry and its effective amplitude, defined below, will be small and hence neglected.

Similar to the adiabatic mode, we can also define the asymmetric tensor and isocurvature perturbations as follows 
\ba
h_{+,\times} &=& h_{+,\times}^0 (1+A_T \,  {\bf  \hat p.\hat n} )
\\
\calS_m &=& \calS_m^0 (1+A_S \,  {\bf  \hat p.\hat n} )
\ea
where $h_{+,\times}^0 $ represent the isotopic tensor perturbations for the two tensorial degrees of freedom $ \{+,\times \} $,  while  $\calS_m^0$ represents the isotopic isocurvature perturbation. As mentioned before, we do not need to introduce any additional scale-dependence for tensor and isocurvature perturbations since they naturally decay quickly for smaller scales due to their non-trivial transfer functions. As a result $A_T$ and $A_S$ are assumed to be scale-independent amplitudes of the dipole modulations for each mode.

Now we can obtain the total $\Cl$'s from the combination of all the above independent modes when ${\bf  \hat p.\hat n} =0$ and  ${\bf  \hat p.\hat n} = 1$  as follows. For the case in which ${\bf  \hat p.\hat n} =0$ we have 
\ba
\label{Cl0}
\Cl^0 &=& \Cl^{\zeta_0} + R_S  \Cl^S + R_T \Cl^T   \quad \quad   ( {\bf  \hat p.\hat n} =0 ) \, ,
\ea
which corresponds to the total power spectrum in the absence of asymmetry (and hence the super script $0$). 
 For the case ${\bf  \hat p.\hat n} =1$ we get 
\ba
\label{Cl}
\Cl^{max} &=& \Cl^{\zeta_0} +  2 A_\zeta \Cl^{\Delta \zeta} +  R_S  (1+A_S)^2 \Cl^S
 \nonumber\\
&+&  R_T (1+A_T)^2 \Cl^T 
\quad  \quad   \quad  \quad   ( {\bf  \hat p.\hat n} =1 ) \, ,
\ea
which is the maximum power spectrum one can get in the asymmetric sky, happening along the preferred direction.
 
In the above relations $\Cl^{\zeta_0}$, $\Cl^S$ and $\Cl^T$ are the adiabatic, isocurvature and tensor angular power spectra, respectively, all normalized by the Planck's best fit value for the adiabatic mode power spectrum in the isotropic limit. Hence, $R_S$ and $R_T$ are defined by 
\ba
R_T \equiv \dfrac{\calP_T}{\calP_\zeta} \quad , \quad 
R_S \equiv \dfrac{\calP_S}{\calP_\zeta}.
\ea
Note that in the above relations we assumed that the adiabatic and isocurvature modes are un-correlated. This simplifies our calculations in this section and is valid for several models of isocurvature perturbations such as axion-type perturbations. In the next section, when we discuss asymmetry from isocurvature modes, we will consider totally correlated and totally anti-correlated cases as well.
Having this said, the interesting point about the above formalism is that $\Cl$'s are formally applicable for all correlations. 

The variables  $R_T$ and $R_S$ are related to the well-known parameters, $r$ (the tensor-to-scalar ratio) and $\beta$ (the fraction of primordial isocurvature perturbations) by \cite{Ade:2013uln}
\ba
r \equiv \dfrac{\calP_T}{\calP_\zeta + \calP_S} = \dfrac{R_T}{1+R_s}
\\
\beta  \equiv \dfrac{\calP_S}{\calP_\zeta + \calP_S} = \dfrac{R_S}{1+R_S}.
\ea
Note that we extended the definition of $r$ to the case in which the isocurvature mode is also present. Also note that for small values of $r$ and $\beta$ they converge to $R_T$ and $R_S$, respectively. 
The Planck constraint, marginalising over all possible correlation angles, is $\beta < 0.075$ \cite{Ade:2013uln}, while it varies for fixed correlations angles or some specific models. For example, for the axion (un-correlated isocurvature) model we have $\beta < 0.036$. To be conservative and make the comparison simpler, we stick to $\beta=0.036$ in most cases but we will consider another case of $\beta=0.075$ as well to observe the dependence of the final results to the value of $\beta$.
As for the parameter $r$, the first Planck release puts the upper bound $r<0.11$ \cite{Ade:2013uln}, while Bicep2 collaboration claimed an observation of $r \simeq 0.2$ \cite{Ade:2014xna}. Even after allowing for the possible role of galactic dust \cite{Adam:2014bub}, it is still interesting to see how a non-zero value of $r$ can affect asymmetry. In this work, whenever we consider asymmetric gravitational wave we mainly set $r=0.1$ while, for comparison,  we also sometimes check the results for $r=0.2$.

From the difference between the correlations in two extreme directions ( ${\bf  \hat p.\hat n} =0$ and  ${\bf  \hat p.\hat n} = 1$)  one can define the scale-dependent asymmetry by
\ba
 \label{defKell}
K_\ell \equiv\dfrac{1}{2} \dfrac{\Delta \Cl}{\Cl^0}
\ea
from which one can see the dependence of the asymmetry on the scale of observation. 
Following \cite{Erickcek:2009at}, and having in mind Eq. \eqref{Cl} together with Eq. \eqref{defKell}, one can also define the effective amplitude of asymmetry via\footnote{Another alternative definition for the effective amplitude of asymmetry is introduced by McDonald \cite{McDonald:2014lea}
$
A(l) \sim \sum_{\ell' =\ell}^{\ell' =\ell_{max}}(2\ell'+1) \Delta C_{\ell'}/2\sum_{\ell' =\ell}^{\ell =\ell_{max}}(2\ell'+1) C_{\ell'}.$ 
While this definition has more direct relation to the methods of data analysis propose in \cite{Hoftuft:2009rq},  our definition has the advantage that $K_\ell$ can be considered as the amplitude of asymmetry at each $\ell$ which in principle can be an observable quantity. In any case, we do not expect much differences between the results of these two definitions.}
\ba 
\label{def_Aeff}
\dfrac{1}{2}\left[(\A+1)^2-1\right] \equiv \sum_2^{\ell_{max}} \dfrac{2\ell+1}{(\ell_{max}-1) (\ell_{max}+3)} K_\ell.
\ea 
with $\ell_{max} = 64$. Note that for each $\ell$ there are $2\ell+1$ independent modes so we weighted the contribution of each $\ell$ to the amplitude of asymmetry by this factor. Again the above relations work for all temperature and/or polarization correlations. For small $\A$ this reduces to 
\ba
\A \simeq \sum_2^{\ell_{max}} \dfrac{2\ell+1}{(\ell_{max}-1) (\ell_{max}+3)} K_\ell.
\ea
which matches to the definition in \cite{Erickcek:2009at}. We will keep the quadratic term in \eqref{def_Aeff} since $\A$ for polarization can be large for some specific cases. 
Note that for the $TT$ correlations we require $\A^{TT} = 0.07$ to match the observations for the asymmetry in temperature correlations. This observational constraint fixes some parameters and then we can investigate what would be the prediction for other correlations. 

From the above definition of $K_\ell$ it is easy to check that in the absence of lensing, i.e. when the $BB$ power spectrum is totally originated from gravitational wave, one has $K_\ell^{BB} = ((1+A_T)^2-1)/2$ which is independent of scale. Hence $K_\ell$ is trivial in this case and we will not plot $K_\ell^{BB}$ in the next section. As another consequence of this observation one has $\A^{BB} = -1+\vert A_T + 1 \vert$, which is justified in the plots in the next section.


\section{Predictions}

In this section we use the formalism developed above to obtain predictions of dipole asymmetry in polarization. We consider two different classes. The first class is when only one mode is responsible for generating the asymmetry while the other modes are either symmetric or negligible in amplitude. The second class is when two independent modes jointly contribute to the dipole asymmetry. The more generic case in which all three types of modes contribute in dipole asymmetry is more complicated and adds more parameters to the analysis which we do not consider for simplicity.  Below we study each class in turn.

\subsection{Asymmetry in one mode}

Here we study the predictions for asymmetry in polarization auto-correlations and cross-correlations, assuming that only one type of mode generates the asymmetry. In what follows we always fix the six parameters in the $\Lambda CDM$ model to the best fit values from Planck data \cite{Ade:2013zuv}. In addition, the spectral index of isocurvature mode is also assumed to be the same as the adiabatic mode for simplicity. In this and next sections   whenever any of sub-dominant modes (i.e. tensor and isocurvature modes) do not contribute to the dipole asymmetry (i.e. their power spectrum is symmetric) we neglect their sub-leading contributions to the symmetric part of the power spectrum as well (i.e. by setting either $r=0$ or $\beta=0$).

\subsubsection{Asymmetry from adiabatic perturbations}
\label{adiabatic-mode}

\begin{figure}
\includegraphics[scale=.45]{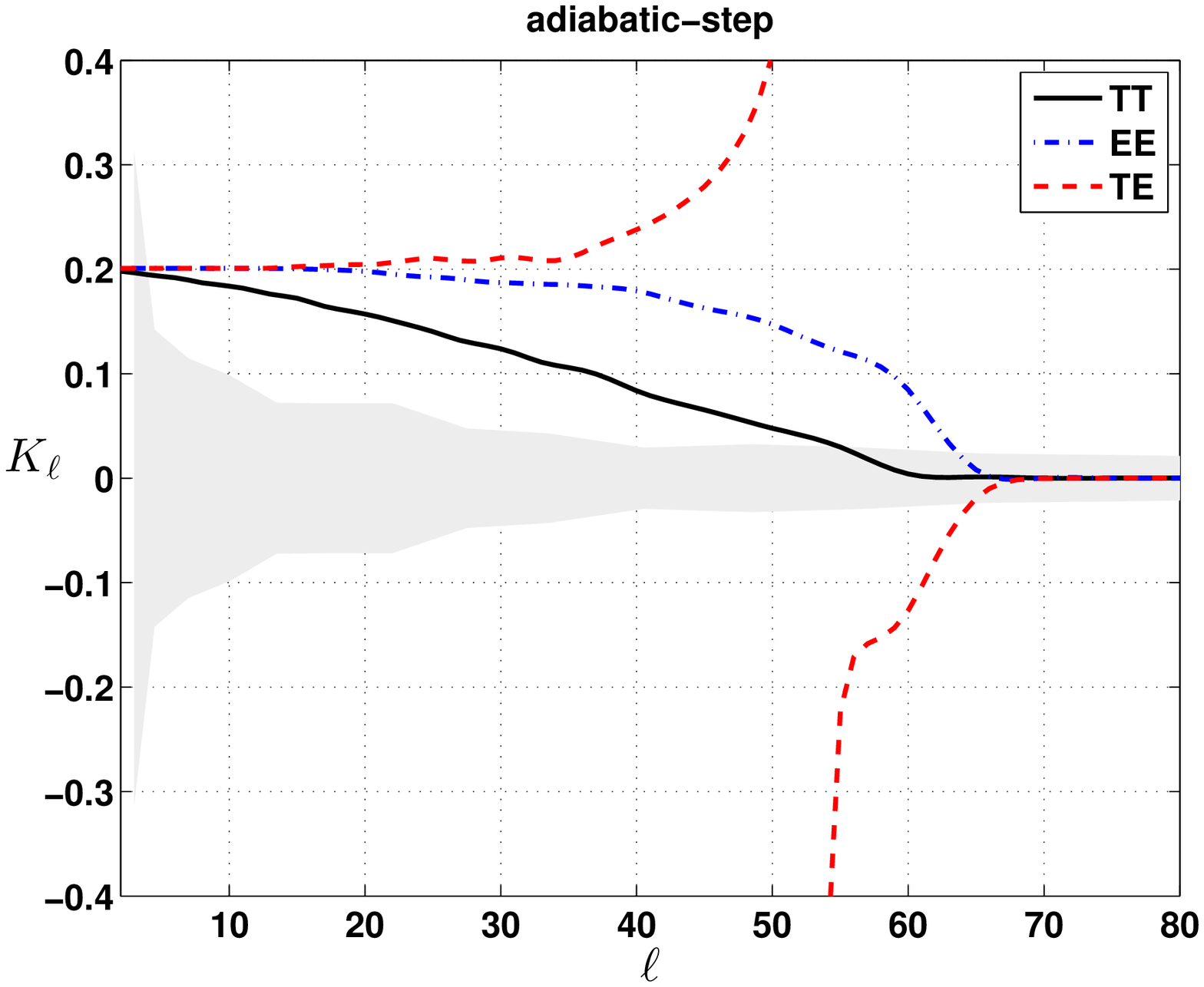}
\includegraphics[scale=.45]{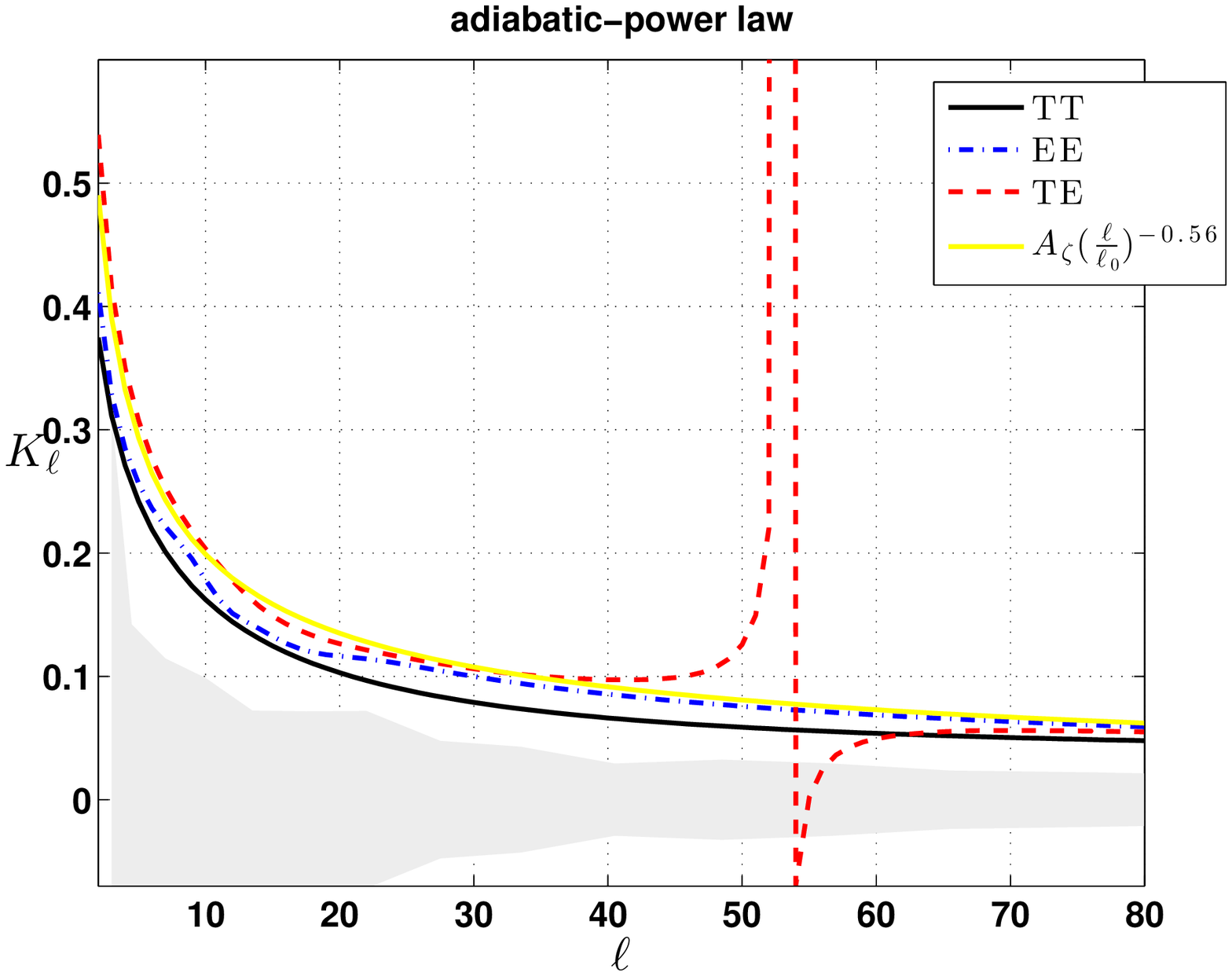}
\caption{The asymmetry $K_\ell$ defined in Eq.~(\ref{defKell}) generated by the adiabatic mode discussed in 
subsection \ref{adiabatic-mode} for different forms of shape function $f(k)$. The top plot is for a step function while the bottom plot is for $ f(k)= (k/k_*)^{\alpha}$ with $\alpha = -0.56$.   The amplitude of asymmetry on $TT$ correlation is fixed to $\A^{TT}=0.07$ for both shapes by tuning the $A_\zeta$. For the case of a step function (top) we have $A_\zeta=0.184$  which yields to $\A^{TE}=1.35$ and $\A^{EE}=0.14$. For the power-law case (bottom) we have $A_\zeta=0.112$,  $\A^{EE}=0.086$ and $\A^{TE}=0.50$. The divergent behavior of $K_\ell^{TE}$ around $\ell = 52$ is because of the fact that $C_\ell^0$ approaches zero at this scale. Note that since there is no
asymmetry from tensor perturbations in this plot, we have $K_\ell^{BB}\simeq 0$, neglecting the lensing effect. The grey shading indicates the cosmic variance  (i.e. $\Delta C_\ell^{var}/2C_\ell$) for the $TT$ correlation in binned multipoles. Note that both sub-dominant symmetric modes are neglected here, i.e. $r=0$ and $\beta=0$ }
\label{adiabatic}
\end{figure}

If the whole asymmetry is generated from adiabatic perturbations, we need a scale-dependent shape function $f(k)$ such that for small scales the amplitude of dipole goes to zero.
To be specific, we consider  two forms for $f(k)$. The first one is a step function with $f(k)=1$ for $k<k_0$ and $f(k)=0$ for $k>k_0$ where we set $k_0=0.004 Mpc^{-1}$ corresponding to an angular multipole $\ell \simeq 64$. The predictions for $TT, EE$ and $TE$ correlations are plotted in top panel of 
Fig. \ref{adiabatic}.  Note that, following our discussions at the beginning of this section, we neglected the contribution to the power spectrum from sub-dominant symmetric modes, i.e. we have set $r=0$ and $\beta=0$.
As it is clear from the plots, despite the fact that the asymmetry is a step function in Fourier space, it behaves very smoothly in $\ell$ space. This is because for each $\ell$ more than one specific $k$ contributes to $\Cl$, though the main contribution comes from the scale which satisfies 
$k\eta_0 =\ell$ ($\eta_0$ is the present value of conformal time).

A more realistic possibility for the scale-dependent asymmetry may be given by the shape function 
 $f(k)= (k/k_*)^{\alpha}$ with $\alpha<0$ in which  $k_*$ is a reference scale. The upper bound on $\alpha$, implied from the lack of dipole asymmetry on quasar scales from the findings of \cite{McDonald:2014lea,McDonald:2014kia},  is $\alpha=-0.56$. We choose this upper bound for studying this shape. Note that since each mode leaves its major imprint on $C_\ell$ with $\ell \simeq k\eta_0$, we expect  $K_\ell$ to behave as $2 A_\zeta(\ell/\ell_*)^{-0.56}$ as a function of $\ell$ where $\ell_*=k_*\eta_0$. This is confirmed by the comparison of this function with the full numerical results in Fig.~\ref{adiabatic}, bottom panel.

As one can check from  Fig.~\ref{adiabatic} the interesting prediction for this case, in which the adiabatic mode generates the entire asymmetry, is that for both shape functions 
the asymmetry on $TE$ and $EE$ correlation are both generically larger than the one for $TT$. To be more precise, we have set $\A^{TT}=0.07$ by tuning free parameter $A_\zeta$ and then predict $\A^{TE}=1.35$ and $\A^{EE}=0.14$ for step function and $\A^{TE}=0.5$ and $\A^{EE}=0.086$ for power law dipole asymmetry.  The divergent behavior of $K_\ell^{TE}$ is because of the fact that $\Cl^0$, which appears in the denominator of $K_\ell$, approaches zero before changing the sign around $\ell=52$. This behavior can be observable in the un-binned data, though it makes the effective amplitude of asymmetry in the $TE$ correlation, in the way we defined here, very sensitive to the parameters. Note that since $K_\ell^{TE}$ is a function of an integer number $\ell$ it is still well defined as $\Cl^0$ for $TE$ never become zero, though it is very small at around $\ell=52$.  Instead of $K_\ell$ one may look at $\Delta C_\ell$ which behaves better but it is not straightforward to relate it to the amplitude of asymmetry. 

 In  the figures presented in  this section we  have plotted the cosmic variance reported by WMAP binned data. That is, we shaded the region obtained by the ratio $\Delta C_\ell^{var}/2C_\ell$ where $\Delta C_\ell^{var}$ is the error due to the cosmic variance. We note that  while this is not the actual error for dipole asymmetry but it is still useful for indicating  how large the error due to the cosmic variance on scale dependent dipole asymmetry $K_\ell$ can be. While the cosmic variance decreases by binning the data, we do not expect much difference between the amplitude of asymmetry in binned and un-binned data. Hence it is much better to look for $K_\ell$ in binned data in future data analysis. This would be somehow similar to the method employed in \cite{Akrami:2014eta} though in that work each scale is considered together with the smaller scales, making the scale dependence of asymmetry unclear. Here, however, we propose to look at the binned data at each scale $\ell$ with binning size $\Delta \ell$ to extract important information regarding the scale dependence of asymmetry. 

\subsubsection{Asymmetry in isocurvature perturbations}

\begin{figure}
\includegraphics[scale=.4]{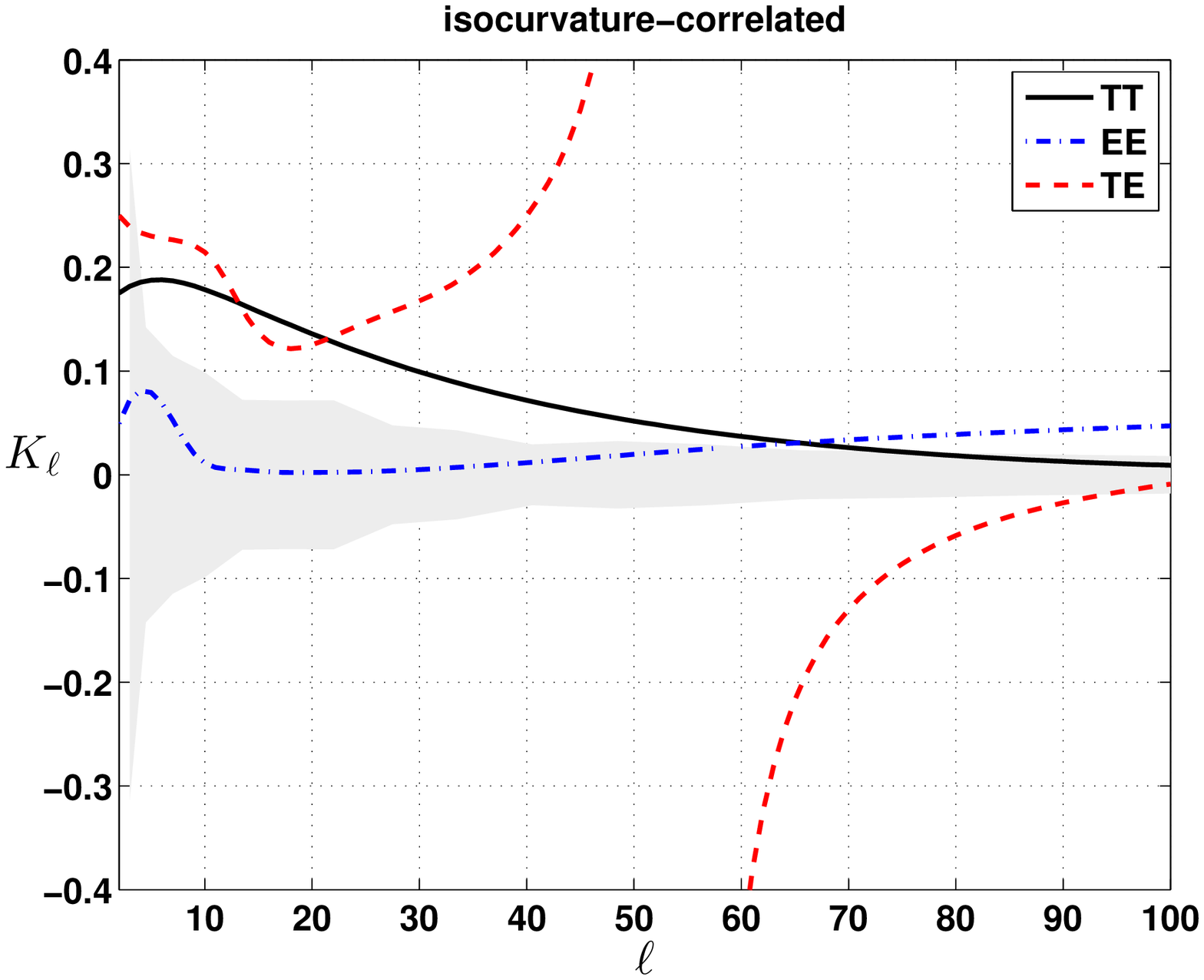}
\includegraphics[scale=.4]{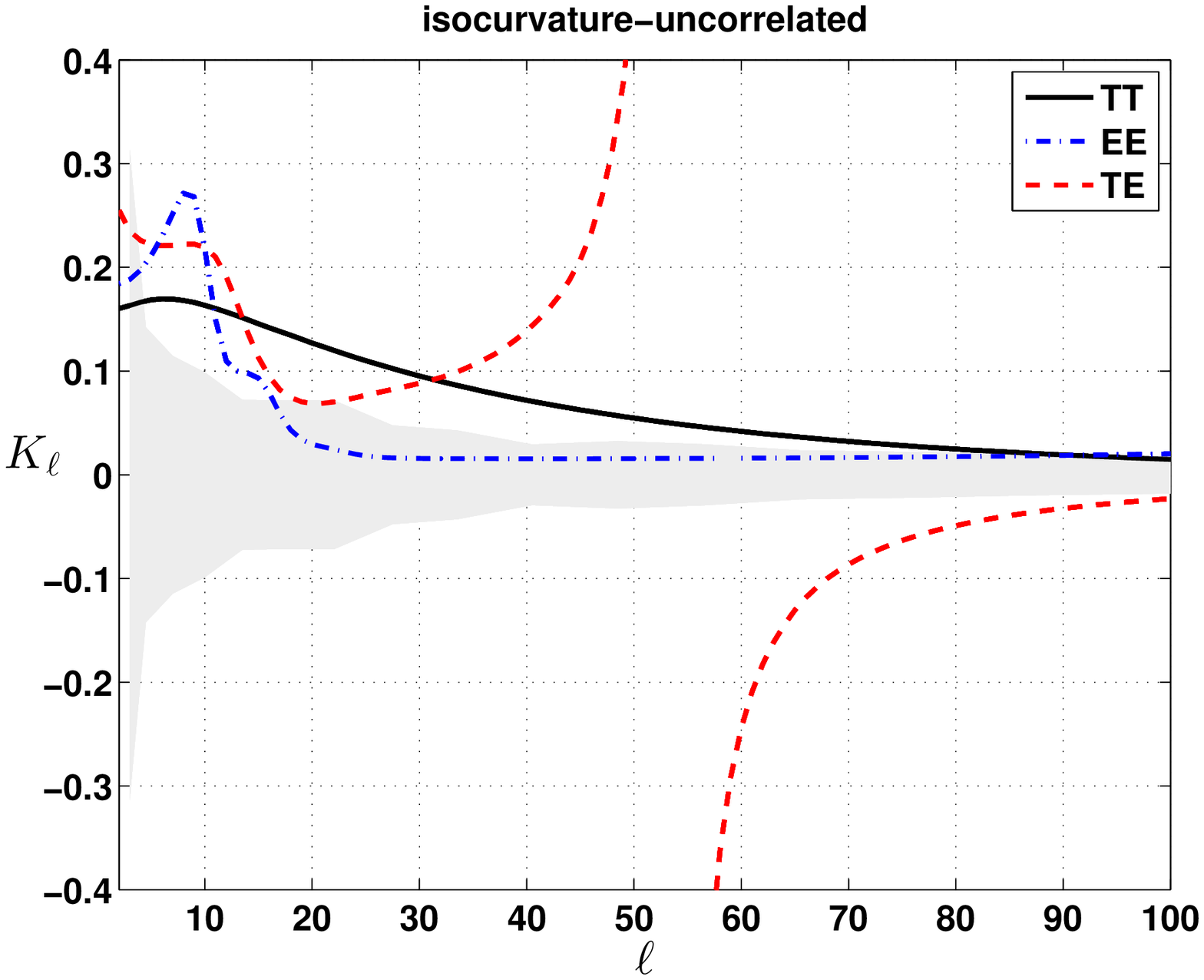}
\includegraphics[scale=.4]{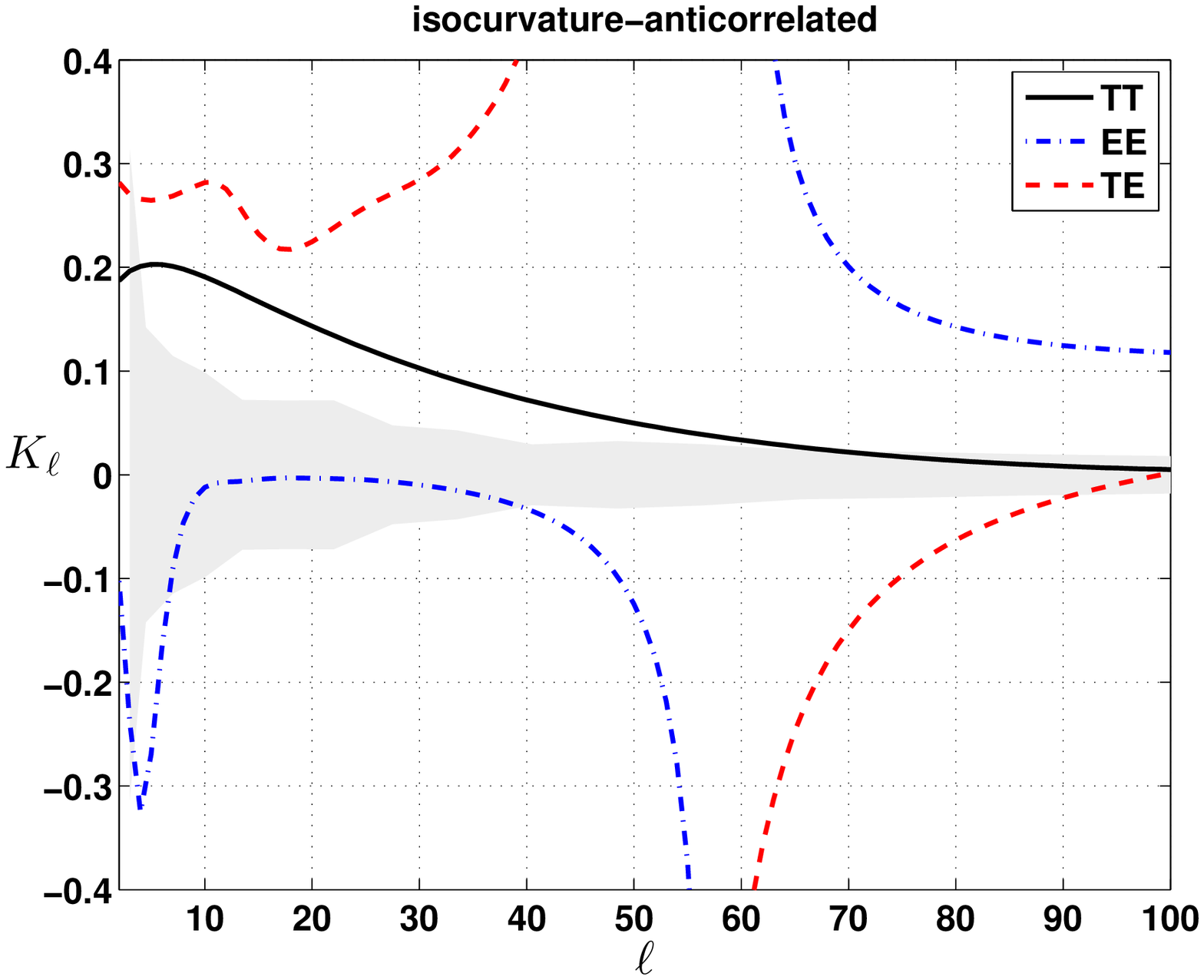}
\caption{The asymmetry $K_\ell$ defined in Eq.~(\ref{defKell}) generated by isocurvature perturbations in the totally correlated (top), un-correlated (middle) and totally anti-correlated (bottom) cases. For all cases we set $\A^{TT}=0.07$ by tuning $A_S$. For the correlated isocurvature (top) we have set $A_S=0.6$ which yields 
$\A^{EE}=0.017$ and $\A^{TE}=0.09$. For the un-correlated case (middle) we have $A_S=1.45$ which gives $\A^{EE}=0.026$ and $\A^{TE}=-0.029$. Finally for the anti-correlated case (bottom) we have $A_S=-1.05$ which results in $\A^{EE}= 0.016$ and $\A^{TE}=-0.317$. For all cases we have set the isocurvature fraction $\beta = 0.036$. Neglecting the lensing effect we also predict $\A^{BB}\simeq 0$ for all cases.
The grey shading indicates the cosmic variance (i.e. $\Delta C_\ell^{var}/2C_\ell$) for the $TT$ correlation in binned multipoles. Note that in all cases the contribution of symmetric tensor mode to the total power spectrum is neglected,  i.e we have $r=0$.}
\label{Kl-iso}
\end{figure}

Now let us consider the case in which asymmetry is generated entirely from the isocurvature mode. We consider three different cases for the correlation between adiabatic and isocurvature modes: totally correlated, totally anti-correlated and un-correlated. Note that Eqs.~(\ref{Cl}) and (\ref{Cl0}) are valid for the un-correlated case only. For the correlated/anti-correlated case we may use the following equations
\ba 
\Cl^{max} = \Cl^{\zeta_0} + R_S (1+A_S)^2 \Cl^{S} \pm \sqrt{R_S} (1+A_S) \Cl^{cor}
\ea 
in which we have ignored the contribution of the tensor mode and the isocurvature mode is assumed to be the only asymmetric mode. In the last term, the plus (minus) sign is for the correlated (anti-correlated) isocurvature mode and $\Cl^{cor}$ is the cross correlation power spectrum normalized to the best fit value of adiabatic amplitude (COBE normalization). Comparing this with the isotropic power spectrum
\ba 
\Cl^0 = \Cl^{\zeta_0} + R_S  \Cl^{S} \pm \sqrt{R_S}  \Cl^{cor} \, ,
\ea 
one can readily obtain $K_\ell$ for which we plotted the results in Fig. \ref{Kl-iso}. For all plots we have set $\beta=0.036$ and kept $\A^{TT}=0.07$ fixed and then predict $\A^{EE}=0.017$ and $\A^{TE}=0.09$ for correlated case,   $\A^{EE}=0.026$ and $\A^{TE}=-0.029$ for the un-correlated case and finally  $\A^{EE}= 0.016$ and $\A^{TE}=-0.317$ for the anti-correlated case. The interesting observation is that for the totally correlated and uncorrelated isocurvature the amplitude of asymmetry is smaller for $EE$ correlation in comparison with that in $TT$ correlation. This is in contrast with what we observed in the case of asymmetric adiabatic mode. An anti-correlated isocurvature can induce larger asymmetry in $EE$ because $\Cl^0$ becomes very small in some regions similar to what happens for $TE$ correlation.

\subsubsection{Asymmetry in tensor perturbations}

\begin{figure}
\includegraphics[scale=.4]{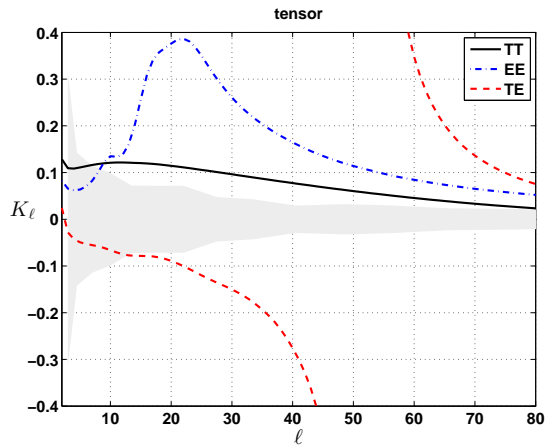}
\caption{The asymmetry $K_\ell$ defined in Eq.~(\ref{defKell}) generated by tensor modes. Here we have $A_T=1.6$ and $r=0.1$ to obtain $\A^{TT}=0.07$ and we predict $\A^{TE}=1.31$, $\A^{EE}=0.15$.  Note that the $K_\ell$ for $BB$ correlation is trivial since only tensor mode contributes to this correlation (after neglecting lensing effect) and hence $K_\ell$ is independent of scale. It is then clear that in this case we have $\A^{BB}=A_T=1.6$. The grey shading indicates the cosmic variance  (i.e. $\Delta C_\ell^{var}/2C_\ell$) for the $TT$ correlation in binned multipoles. Note that here we have set $\beta=0$.}
\label{Kl-tensor}
\end{figure}

One can also consider the case when the asymmetry is generated entirely from tensor modes. The results are shown in Fig.\ref{Kl-tensor}. Note that in this plot we neglected the contribution from isocurvature mode as it has a sub-dominant amplitude and also does not contribute to the asymmetric part. Here we have set $r=0.1$ and tuned $A_T$ such that we recover 
$\A^{TT}=0.07$ and then we predict $\A^{TE}=1.31$, $\A^{EE}=0.15$ and $\A^{BB}=1.6$.
Note that because the tensor power spectrum is already suppressed relative to the adiabatic power, and there is an upper bound on $r$, we need a large value for $A_T$ to obtain the observed dipole asymmetry in the $TT$ correlation. The interesting consequence of this fact is that the asymmetry in the $BB$ correlation becomes very large, as it is independent of the value of $r$. Hence if the asymmetry comes solely from tensor perturbations, one predicts a large asymmetry in the $BB$ correlation. 

Note that asymmetry in other correlations of polarization ($EE$ and $TE$) are also larger than that in the $TT$ correlation, hence there is no degeneracy between the predictions of isocurvature and tensor perturbations, if only one mode is responsible for asymmetry. However, if there is more than one asymmetric mode, the situation is much more complicated. We will study this possibility in some detail in the next section. 

\subsection{Asymmetry in two different modes}
\begin{figure}
\includegraphics[scale=.34]{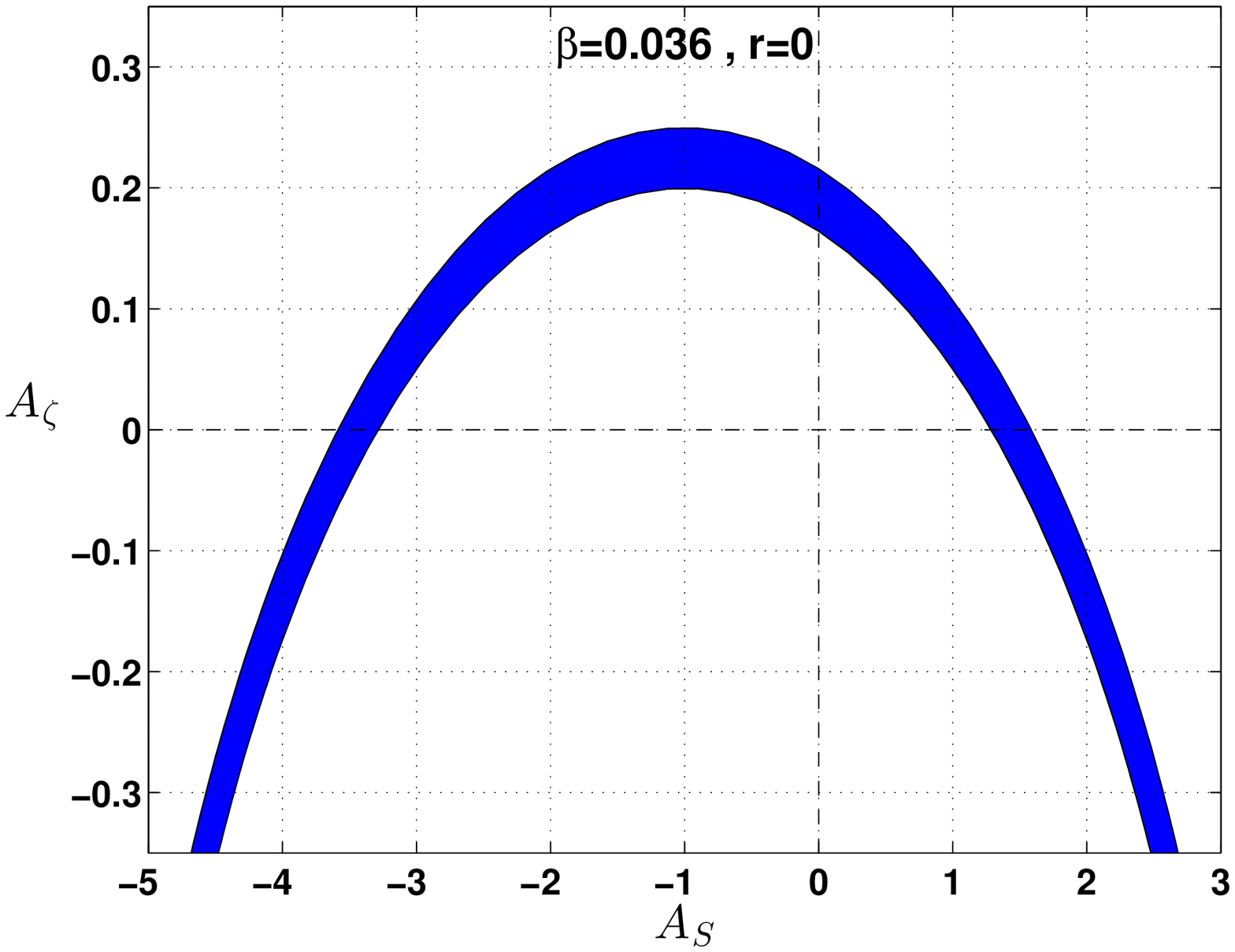}
\includegraphics[scale=.34]{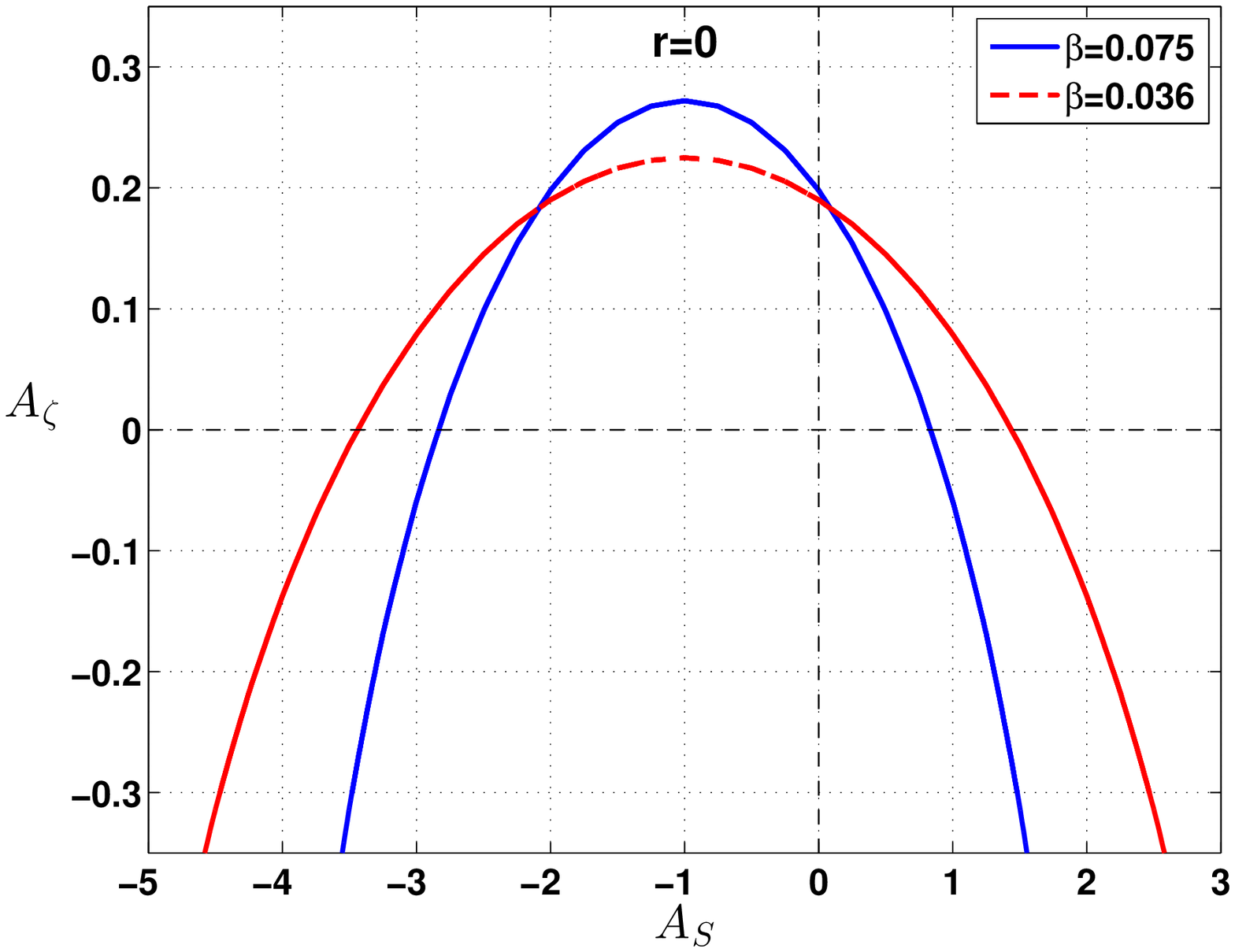}
\\
\includegraphics[scale=.34]{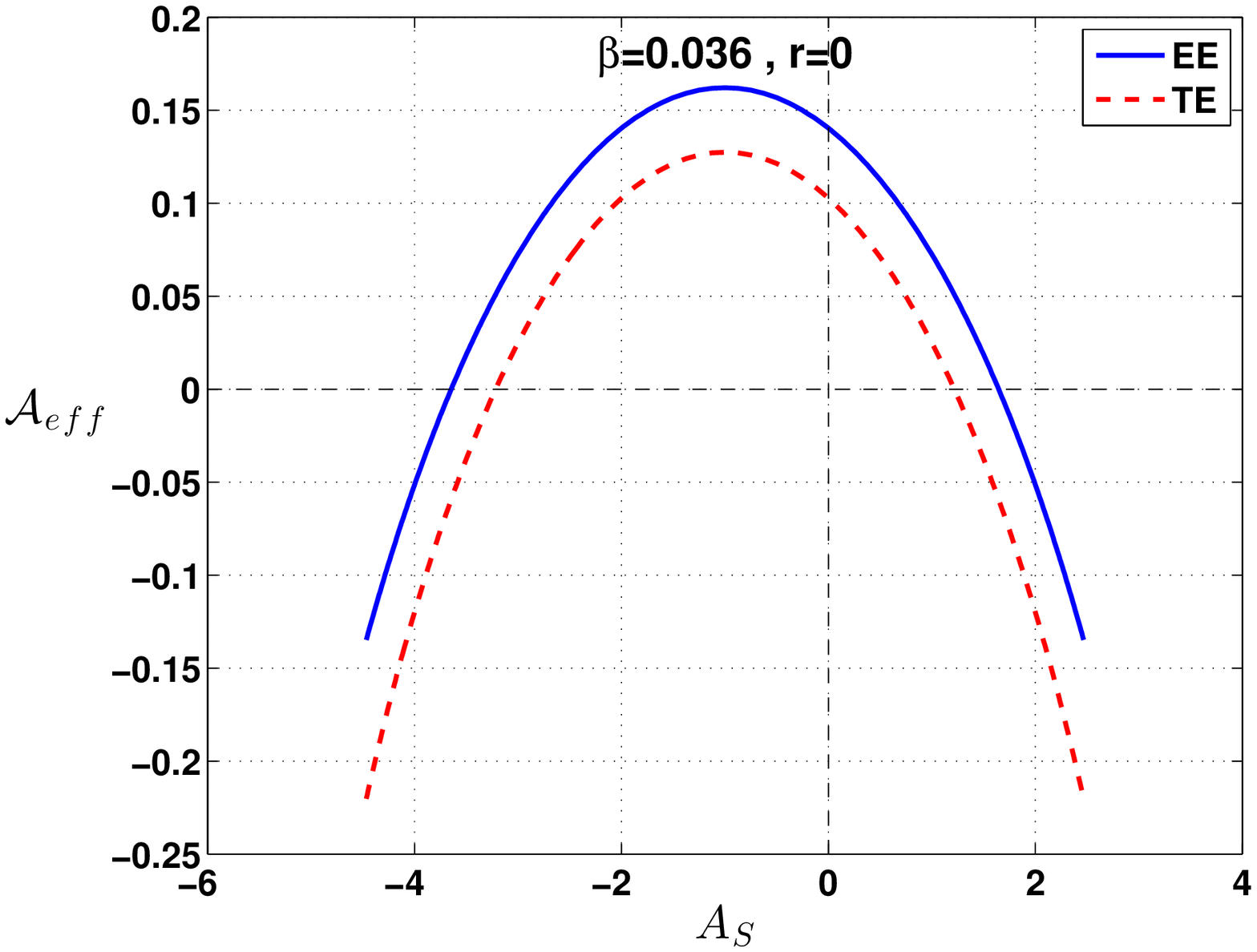}
\includegraphics[scale=.34]{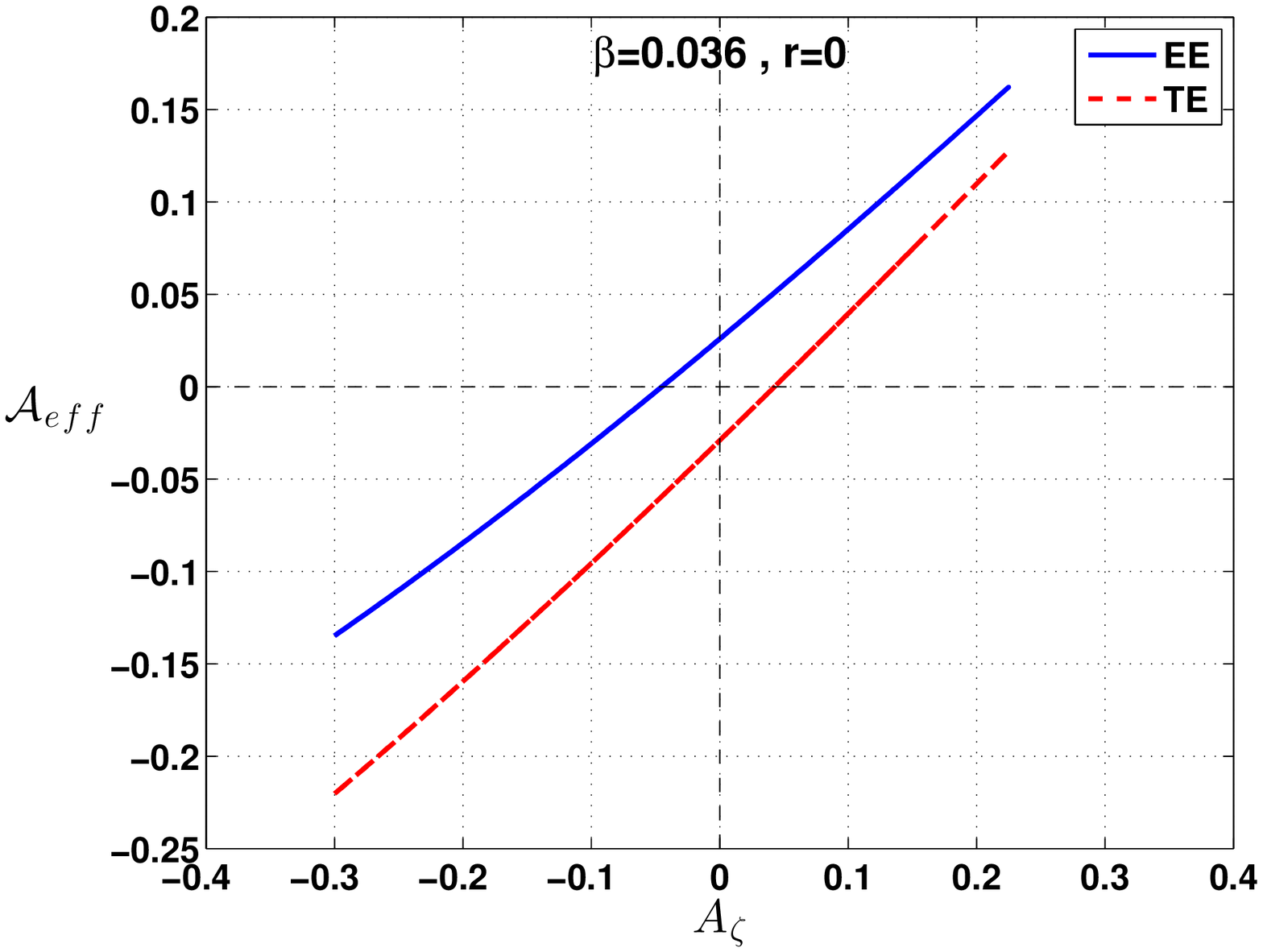}
\caption{Plots for the case in which we turn on both $A_\zeta$ and $A_S$. Top: Contour plot for $A_\zeta$ vs. $A_S$ in which $\A^{TT}$ remains within the $3.5 \sigma$ region of observation, i.e. $\A^{TT}=0.07 \pm 0.01$. The second plot from the top shows how $A_\zeta$ and $A_S$ have to change simultaneously to keep $\A^{TT}=0.07$  unchanged. Fixing $\A^{TT}$ by varying both $A_\zeta$ and $A_S$, the two bottom plots represent the asymmetry prediction for polarization. For the top contour plot and the two lower plots we set $\beta=0.036$ while in the second plot from the top
we have presented the predictions for $\beta=0.075$ and $\beta = 0.036$. Note that in these plots we have set $r=0$, i.e. no gravitational wave is present. As a result, neglecting lensing effect, we have  $\A^{BB} \simeq 0$.}
\label{zeta-s}
\end{figure}

\begin{figure}
\includegraphics[scale=.35]{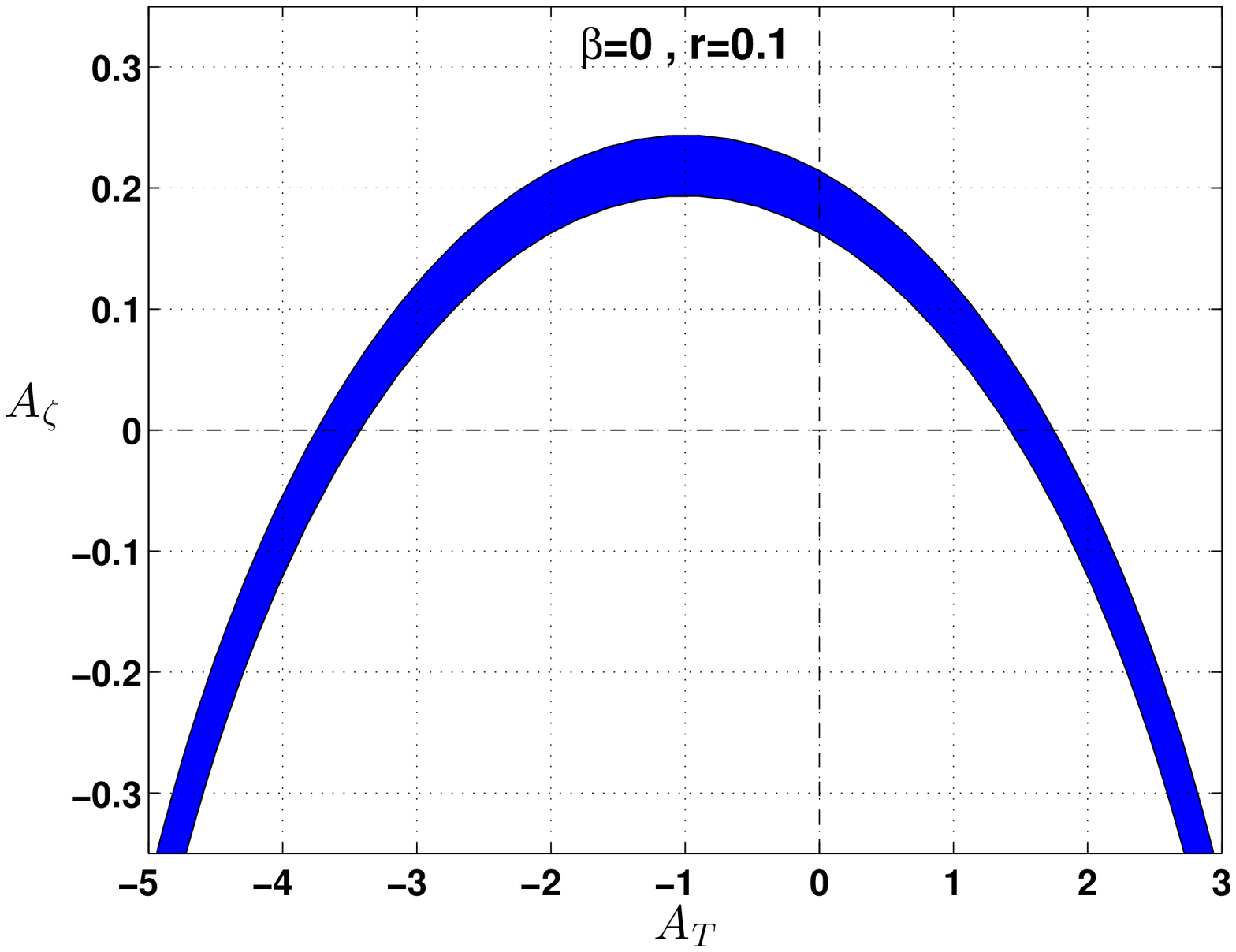}
\includegraphics[scale=.35]{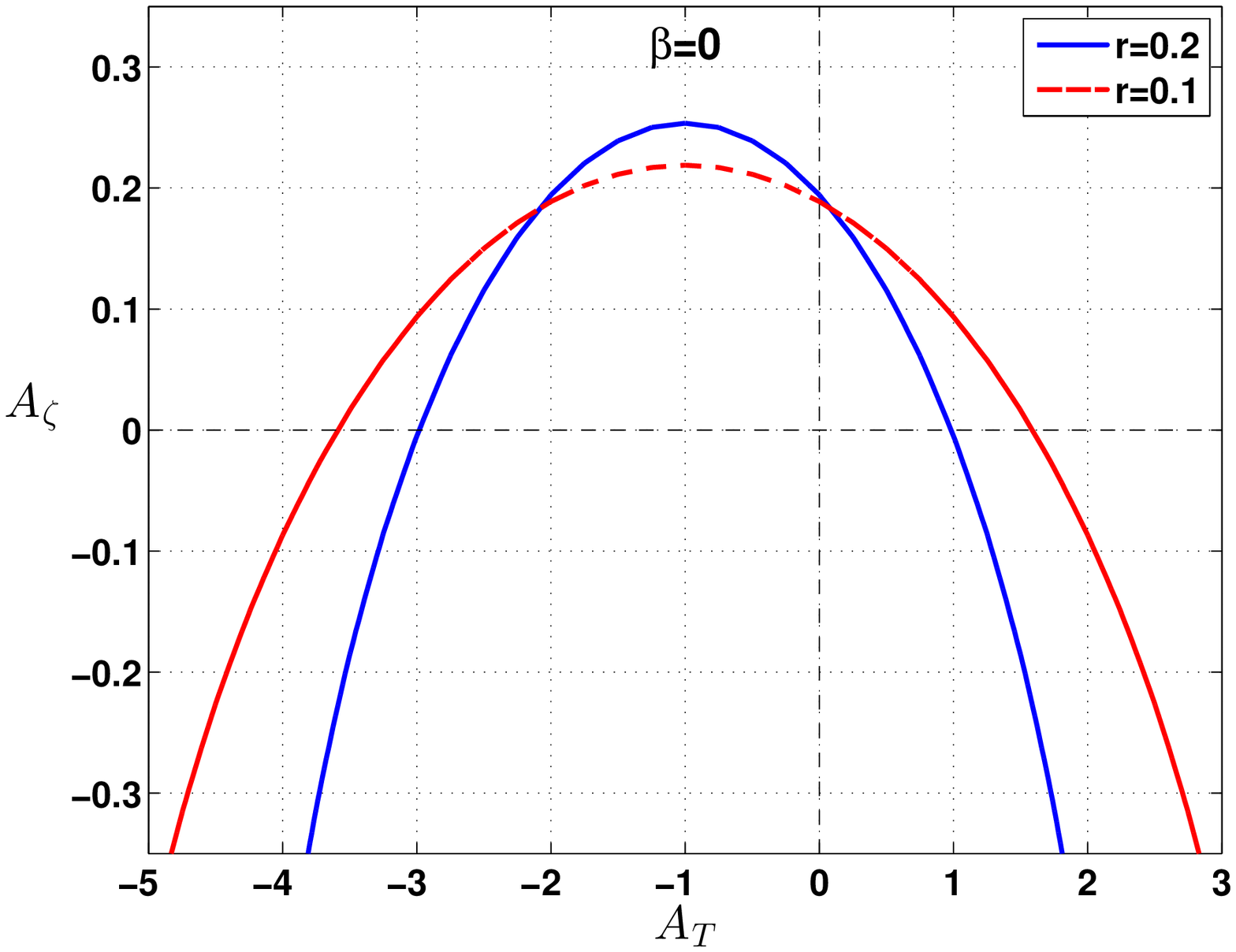}
\includegraphics[scale=.35]{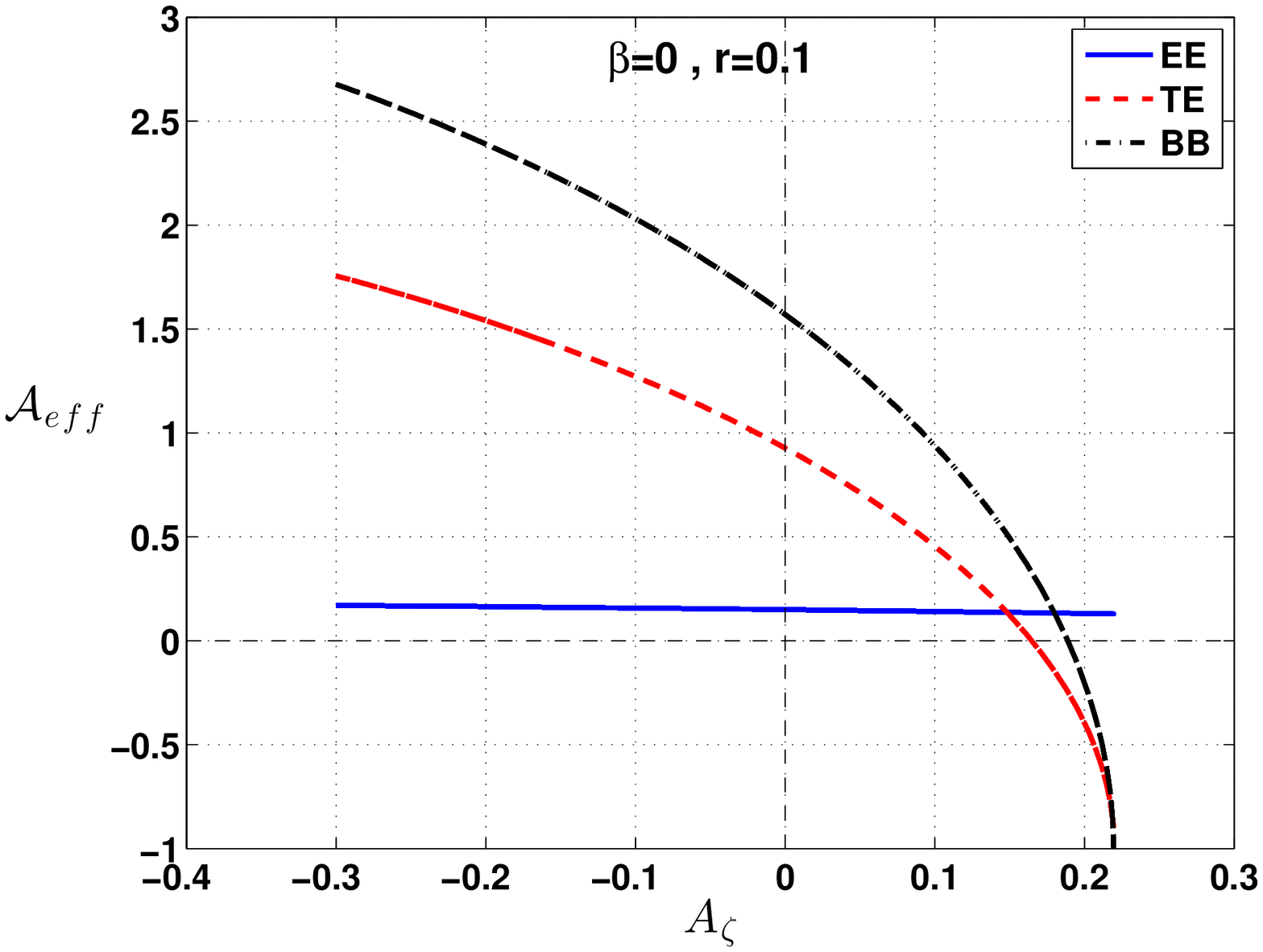}
\includegraphics[scale=.35]{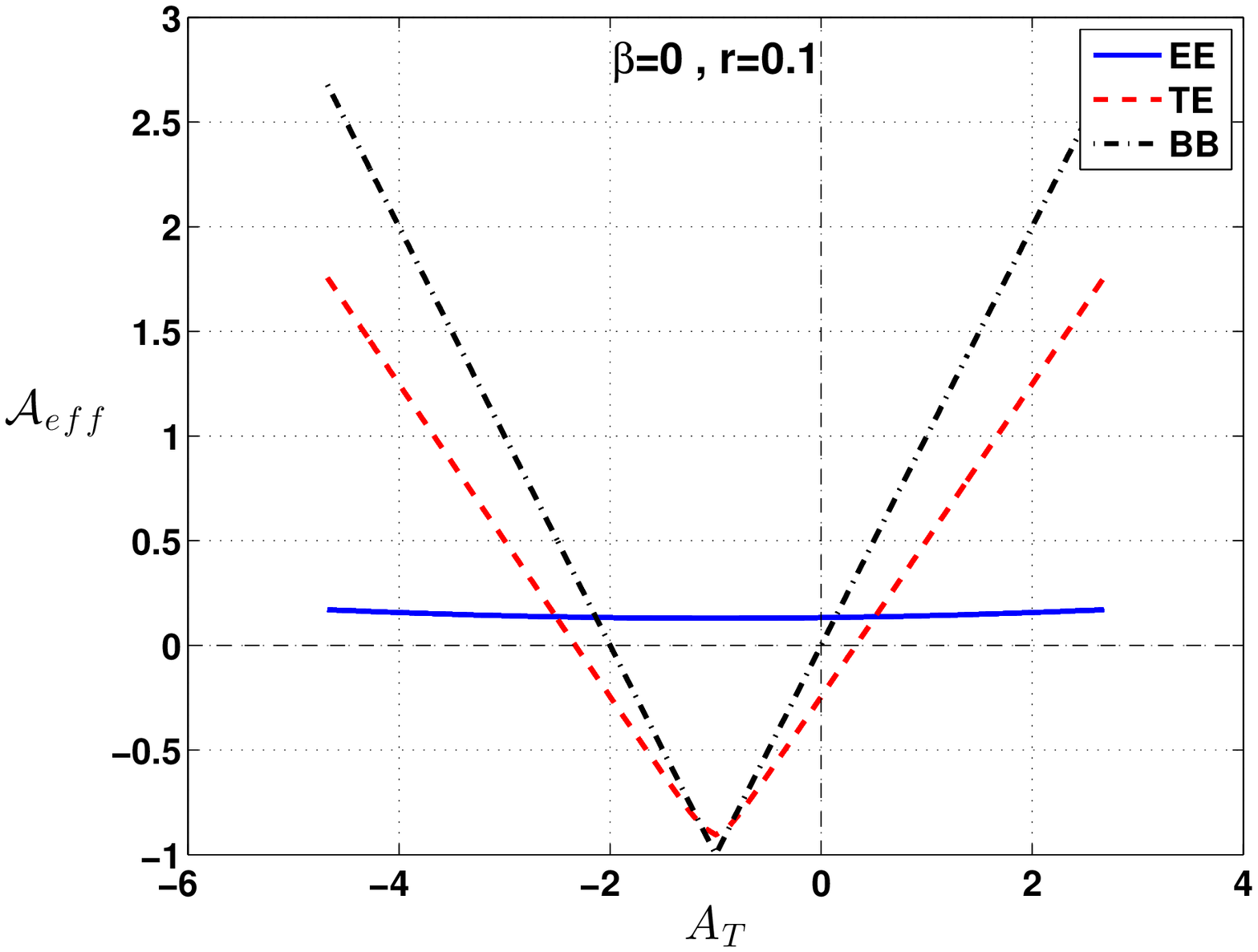}
\caption{ The same as in Fig. \ref{zeta-s} but for the case with asymmetric tensor and adiabatic modes. Top: Contour plot for $A_\zeta$ vs. $A_S$ in which $\A^{TT}$ remains within the $3.5 \sigma$ region of observation, i.e. $\A^{TT}=0.07 \pm 0.01$. The second plot from the top shows how $A_\zeta$ and $A_T$ have to change simultaneously to keep $\A^{TT}=0.07$  unchanged.  The two bottom plots then represent the prediction for 
asymmetry as both $A_\zeta$ and $A_T$ change while keeping $\A^{TT}$ fixed. For the top contour plot and the two lower bottom plots we have set $r=0.1$ while in the second plot from the top we have presented the predictions for
$r=0.2$ and $r=0.1$. }
\label{zeta-t}
\end{figure}

\begin{figure}
\includegraphics[scale=.33]{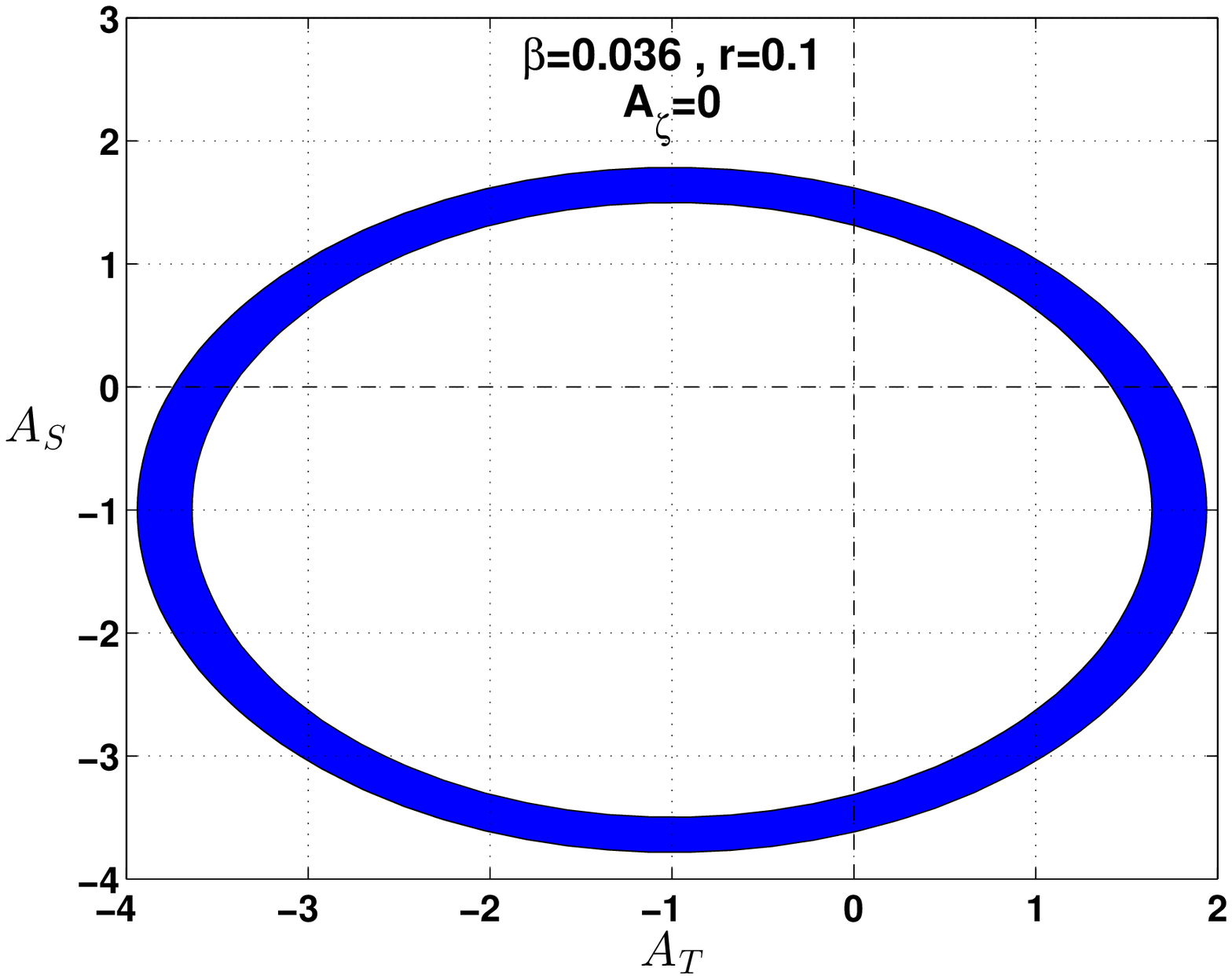}
\includegraphics[scale=.33]{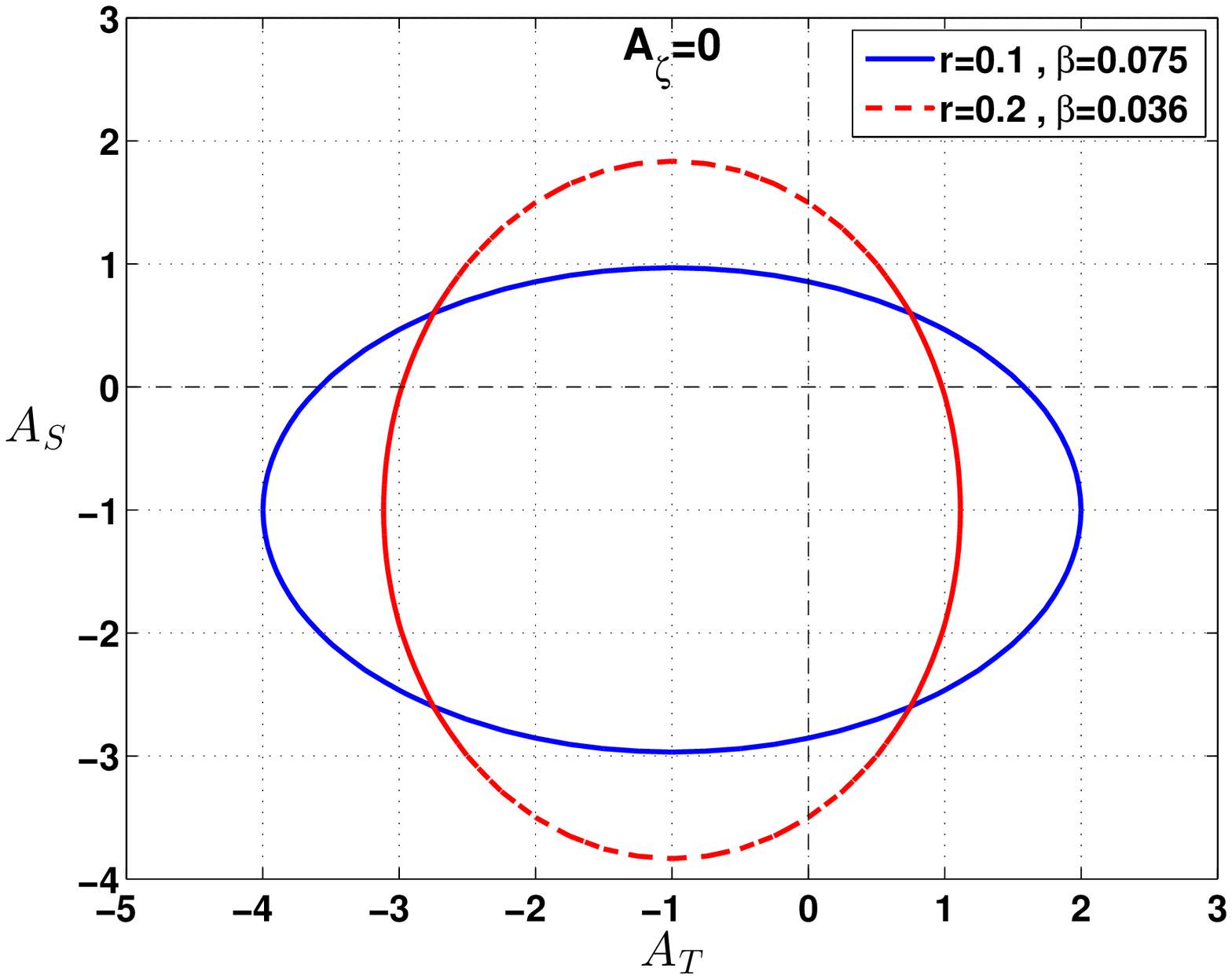}
\\
\includegraphics[scale=.33]{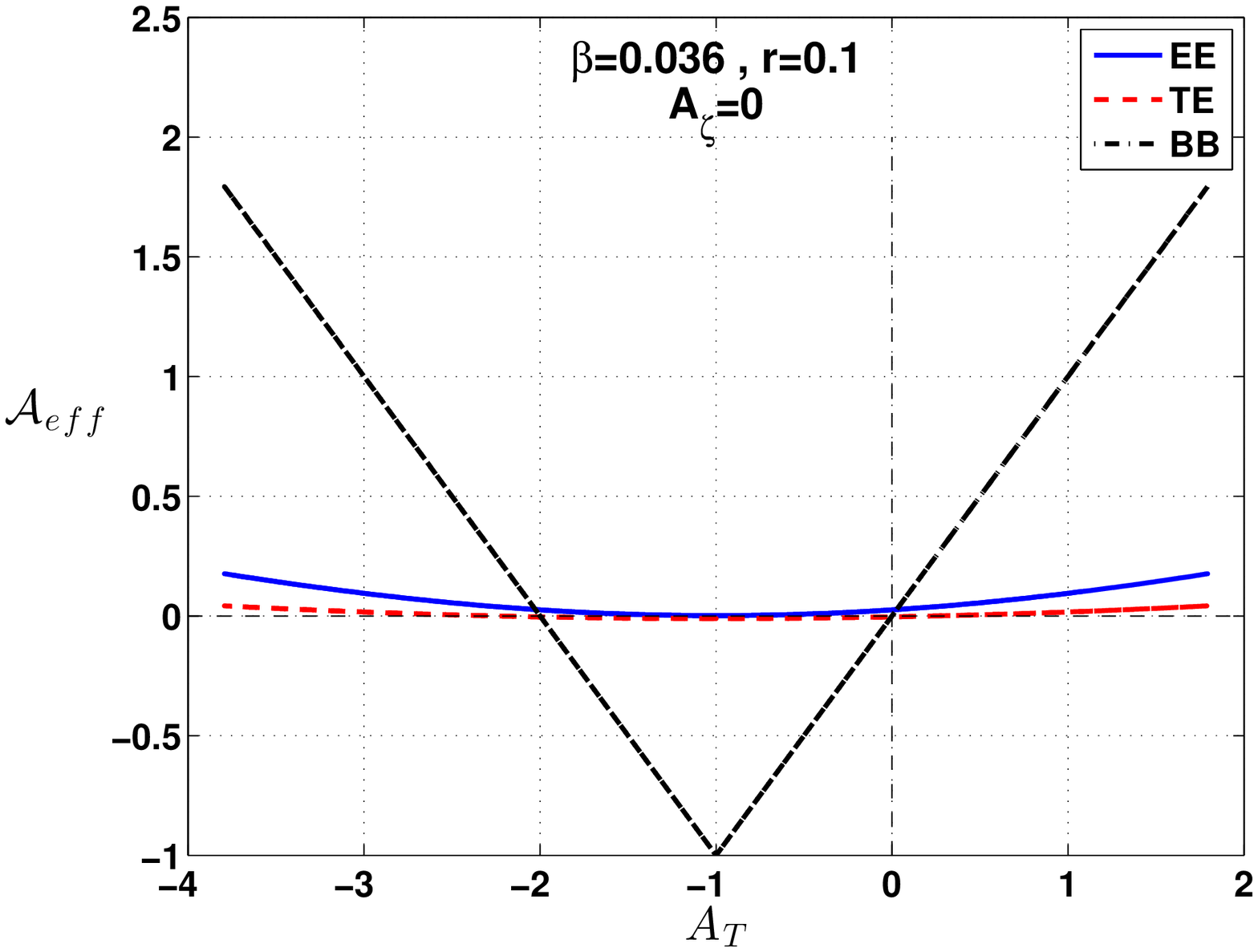}
\includegraphics[scale=.33]{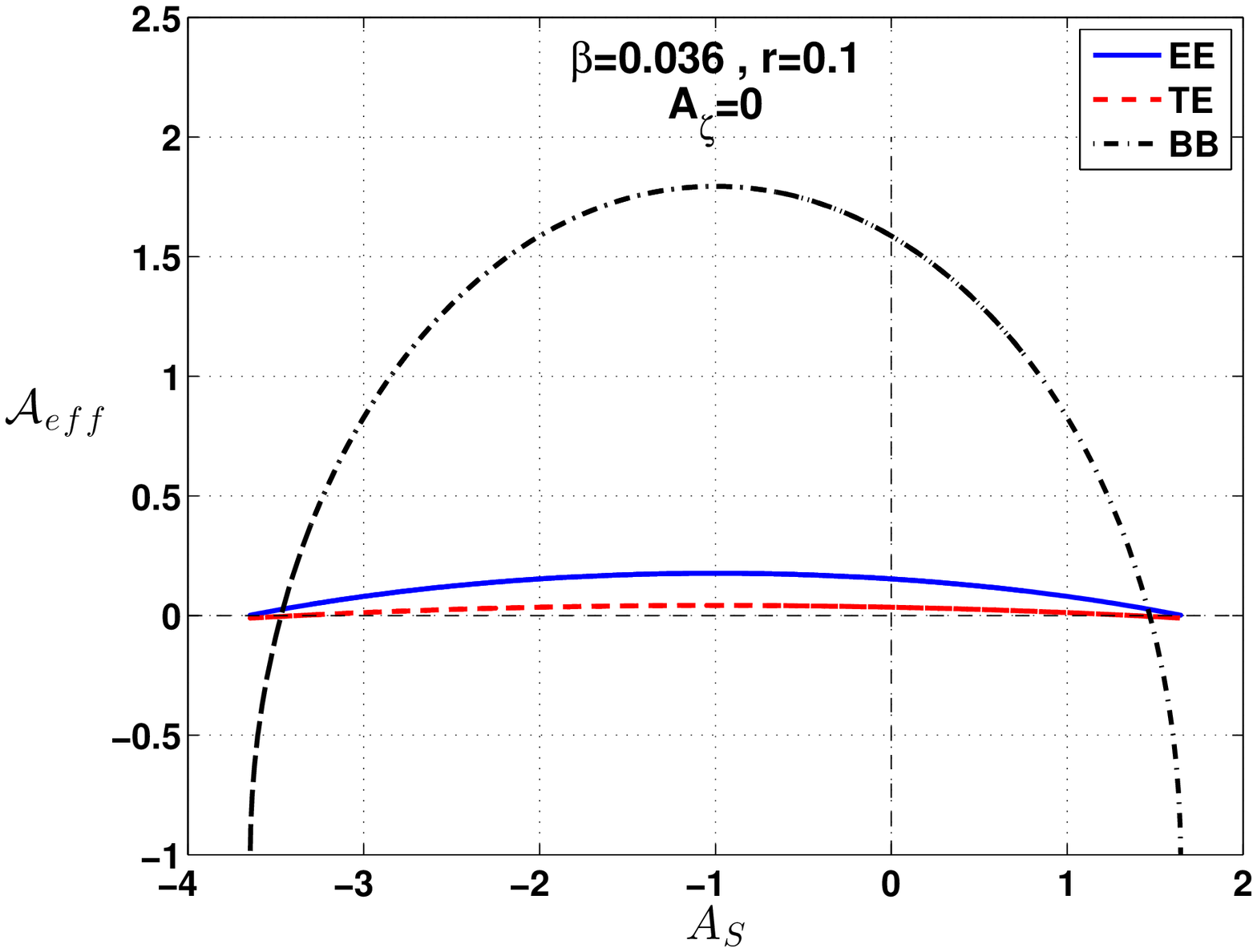}
\caption{ The same as in Fig. \ref{zeta-s} but for the case in which $A_S$ and $A_T$ are non-zero while the adiabatic mode is symmetric. Top: Contour plot for $A_S$ vs. $A_T$ for the $3.5 \sigma$ region of observation, i.e. $\A^{TT}=0.07 \pm 0.01$. The second plot from the top shows the simultaneous change in $A_T$ and $A_S$ keeping $\A^{TT}=0.07$  fixed.  The two bottom plots then represent the asymmetry prediction as both $A_T$ and $A_S$ change keeping $\A^{TT}$ fixed. For the top contour plot and the two lower bottom plots we have set $r=0.1$ and $\beta=0.036$ while in the second plot from the top we have presented the predictions for the set of parameters $(r=0.2, \beta = 0.036)$ and $(r =0.1, \beta =0.075)$. Note that the adiabatic mode is assumed to be symmetric, i.e. $A_\zeta=0$.}
\label{s-t}
\end{figure}

In this section, we study the predictions of asymmetry in polarization in the case when two different modes jointly generate the asymmetry. The results for asymmetry generated in polarization correlations
by $(A_\zeta, A_S)$, $(A_\zeta, A_T) $ and $(A_S, A_T) $ are respectively shown in Figs. ~\ref{zeta-s}, \ref{zeta-t} and \ref{s-t}. 

In these plots we have restricted $A_\zeta$ to a  small value because of the tight observational constraints, while $A_S$ or $A_T$ are allowed to be arbitrarily large. We fix the parameters regarding the amplitude of each power spectra (e.g. $r$ and $\beta$) to the values that satisfy the observational constraints. Then we allow the parameters in the asymmetric parts (e.g. $A_S$, $A_T$) to vary while we keep  $\A^{TT}=0.07$ fixed (except for the contour plots) and then calculate the asymmetry in polarization for each set of parameters that satisfies the latter condition. Note that in this section we only consider a step function for the scale-dependent asymmetry in adiabatic mode as we do not expect much qualitative difference between power law and step function adiabatic dipole asymmetry. We also restrict ourselves to the case of un-correlated isocurvature perturbations for simplicity. 
As for the contours in the top plots of each figure we shaded the region consistent with the tightest observational constraint for dipole asymmetry, i.e. $\A^{TT} = 0.07 \pm 0.01$ with $3.5 \sigma$ accuracy. Again, note that as long as each sub-dominant mode is symmetric its contribution to the total power spectrum is neglected. 

As is clear from the plots, the amplitude of asymmetry in polarization can vary depending on the values of the parameters and we somehow lose the predictability. Obviously the reason is that the number of free parameters is larger than for a single asymmetric perturbation and $\A^{TT}$ fixes only one parameter. However,  it will be very interesting to perform a detailed analysis with the upcoming data allowing all the relevant asymmetry parameters free to vary.


\section{Conclusions and summary of results}

In this work we have explored the predictions for the asymmetric polarization correlations and cross-correlations $TE, EE$ and $BB$  assuming the dipole asymmetry observed in the $TT$ angular power spectrum on large scales has a primordial origin. We have allowed for adiabatic, matter isocurvature and tensor perturbations to be responsible for the asymmetric part of the CMB anisotropies. We have followed a phenomenological approach without specifying a physical model for the origin of the asymmetry in each primordial perturbation.  We also allowed for the case in which more than one
 mode, i.e. adiabatic-isocurvature, adiabatic-tensor and isocurvature-tensor,  jointly 
contribute to the asymmetry. We have fixed the amplitude of the $TT$ power asymmetry to be $\A^{TT}=0.07$ to fit the current Planck constraint on temperature power asymmetry on large angular scales and then investigated the predictions for polarization correlations. 

Here let us summarize the main results and the generic picture which emerges from our analysis. Note that the predictions we outline below are rather robust against changing the parameters such as $\beta$ and $r$.

\begin{itemize}

\item
 If the asymmetry originates from either adiabatic or tensor perturbations then we find that the fractional asymmetry is larger in the E-mode polarisation, $\A^{EE} > \A^{TT}$, while we find $\A^{EE} < \A^{TT}$ if the asymmetry comes from totally correlated or un-correlated isocurvature perturbations. This observation is not necessarily true for anti-correlated isocurvature perturbations as in this case $\A^{EE}$ is sensitive to the free parameters especially due to the divergent behavior.

\item  $\A^{TE}$ is very sensitive to parameters such as $r$ and $\beta$ . This is because $K_\ell^{TE}$ diverges around $\ell \simeq 52$ where $\Cl^{TE}$ approaches to zero. So just looking at  $\A^{TE}$ is not so conclusive and instead one can extract information from the scale dependence of asymmetry in $TE$ correlation, i.e. $K_\ell^{TE}$.


\item 
Neglecting the lensing effect, the asymmetry in $BB$ polarization would be zero if either of scalar modes are responsible for asymmetry.

\item In order to obtain $\A^{TT} \simeq 0.07$ from either tensor or isocurvature perturbations we require very large asymmetry amplitudes for these perturbations. Roughly speaking we require $\Delta \calP^{\calS,T}/\calP^{\calS,T} \gtrsim 1$. This is because any feature due to tensor or isocurvature perturbations is already constrained by observational bounds on $r$ and $\beta$. The important consequence of this is that if an asymmetric tensor mode has an observable effect on the $TT$ correlation then the asymmetry in the $BB$ correlation has to be large since it is independent of $r$.

 \item Without loss of generality we take $\A^{TT}>0$. However other correlations can have negative sign for their asymmetry with respect to $\A^{TT}$. The relative sign is physically meaningful, showing that the asymmetry in different correlations can be in opposite directions.

\item 
If there is more than one perturbation mode contributing to the asymmetry (adiabatic, isocurvature or tensor) then their individual contribution would not be distinguishable by just looking at the amplitude of asymmetry. The scale-dependence of the asymmetry then becomes an important discriminator.

\end{itemize}

The above observations are very interesting, but we note that the method we employed to obtain the amplitude of the asymmetry provides a rough estimate. In this work we outlined what one can learn from dipole asymmetry on polarization about the origin of dipole asymmetry. While a more accurate analysis would be needed in order to compare these predictions against forthcoming data we note that the above qualitative observations are robust and can be checked by future data, shedding light on the origin of dipole asymmetry.   \\


{\bf Acknowledgements:} We would like to thank Y. Akrami, X. Chen, R. Emami and Moslem Zarei
for useful discussions and correspondences. MHN is supported in part by a NSF grant PHY-1417421.  
DW was supported by STFC grant numbers ST/K00090X/1 and ST/L005573/1.


{}

\end{document}